# Have Learning Analytics Dashboards Lived Up to the Hype? A Systematic Review of Impact on Students' Achievement, Motivation, Participation and Attitude

Rogers Kaliisa
University of Oslo,
Norway, rogers.kaliisa@iped.uio.no

Kamila Misiejuk
University of Bergen,
kamila.misiejuk@uib.no

Sonsoles López-Pernas
University of Eastern Finland,
sonsoles.lopez@uef.fi

Mohammad Khalil
University of Bergen,
mohammad.khalil@uib.no

Mohammed Saqr
University of Eastern Finland,
mohammed.saqr@uef.fi

### Abstract

While learning analytics dashboards (LADs) are the most common form of LA intervention, there is limited evidence regarding their impact on students' learning outcomes. This systematic review synthesizes the findings of 38 research studies to investigate the impact of LADs on students' learning outcomes, encompassing achievement, participation, motivation, and attitudes. As we currently stand, there is no evidence to support the conclusion that LADs have lived up to the promise of improving academic achievement. Most studies reported negligible or small effects, with limited evidence from well-powered controlled experiments. Many studies merely compared users and non-users of LADs, confounding the dashboard effect with student engagement levels. Similarly, the impact of LADs on motivation and attitudes appeared modest, with only a few exceptions demonstrating significant effects. Small sample sizes in these studies highlight the need for larger-scale investigations to validate these findings. Notably, LADs showed a relatively substantial impact on student participation. Several studies reported medium to large effect sizes, suggesting that LADs can promote engagement and interaction in online learning environments. However, methodological shortcomings, such as reliance on traditional evaluation methods, self-selection bias, the assumption that access equates to usage, and a lack of standardized assessment tools, emerged as recurring issues. To advance the research line for LADs, researchers should use rigorous assessment methods and establish clear standards for evaluating learning constructs. Such efforts will advance our understanding of the potential of LADs to enhance learning outcomes and provide valuable insights for educators and researchers alike.

**Keywords:** Learning analytics dashboards (LADs), systematic review, impact, learning outcomes

# 1 INTRODUCTION

The field of Learning Analytics (LA) has emerged as a promising avenue for leveraging data-driven insights to enhance educational processes and outcomes. LA can provide students and teachers with valuable feedback and support, enabling them to make informed decisions and optimise their learning and teaching practices. However, despite the proliferation of studies within the LA domain, a critical gap remains in understanding the impact and effectiveness of LA interventions, specifically focusing on LA dashboards (LADs) [1].

LADs may be defined as "displays that aggregate different indicators about learner(s), learning process(es) and/or learning context(s) into one or multiple visualisations" [2]p. 37, that have the potential to empower students and teachers by offering valuable insights into their learning and teaching processes [3,4,5]. These interactive tools aim to visualise data and provide actionable information, enabling learners and educators to monitor progress, identify areas of improvement, and make data-informed decisions. Despite over a decade of advancements and innovations within the LA field, there is a dearth of compelling evidence demonstrating the effectiveness and impact of LA interventions [1,6], with only a few individual studies yet reporting mixed results mostly based on small samples [7]. This lack of empirical evidence poses a significant challenge when attempting to justify investments in expensive LA infrastructure and the necessary human resource training.

To bridge this critical gap in our understanding and provide evidence-based insights into the impact of LADs, we performed a systematic and quantitative analysis of studies that have reported on the implementation and effects of LADs. Our quantitative analysis offers a powerful methodological approach to synthesize and quantitatively analyse findings from multiple studies, enabling researchers to draw robust conclusions and identify patterns and trends that may not be evident within individual studies alone [8]. By analysing the effect sizes from studies conducted under different circumstances (e.g., study settings and disciplines), a quantitative review leads to more precise estimates of the true effect sizes and testing of moderator variables [9].

To the best of our knowledge, few studies have conducted a quantitative review with a primary focus on evaluating the impact of LADs on students' learning outcomes so far. The closest study to ours was conducted by [1], who analysed 522 papers published in the LAK conference proceedings and the Journal of Learning Analytics to evaluate how LA has impacted the understanding of learning and provided insights that have been translated into mainstream practice. The authors concluded that while LA research has developed in the areas of focus and analytical approaches, the impact on practice and theory remains limited. While these findings shed light on the impact of LA, the authors did not report the actual effect of the different LA interventions (e.g., dashboards) on students' learning outcomes. [19] conducted a systematic literature review on applying LA to enhance self-regulated learning (SRL). While this study attempted to analyse the effect of LA interventions, the focus was only on SRL.

This study identifies the effect size of individual dashboard studies on different aspects of students' learning outcomes to address the general concerns about their effectiveness and to facilitate further research and application. Through a systematic and quantitative examination of studies reporting the impact of LADs, we can address the current dearth of evidence and provide stakeholders, including researchers, educational institutions, administrators, policymakers, and funding agencies, with an understanding of the benefits and potential challenges associated with the implementation of LA interventions.

# 2 BACKGROUND

## 2.1 Evaluating the impact of learning analytics

According to [10], the impact of LA interventions can be assessed based on four propositions: 1) whether they support learning outcomes, 2) support teaching, 3) deployed widely, and 4) whether they are used ethically. With the view that the primary goal of LA is to support learning [11] in this study, we explore the impact of LADs-arguably the main intervention point for LA, based on one particular proposition (i.e., whether they support students' learning outcomes).

LA, as a multidisciplinary field of research, embraces methods and approaches from disciplines such as data mining, computer science, machine learning, natural language processing, and human-computer interaction [12]. The multidisciplinary nature of the LA field implies that different researchers conceptualise and measure the impact of LA on students' learning outcomes differently. For instance, researchers rooted in the sociocultural perspective tend to focus on analysing classroom dialogue (both written and spoken) with the view that knowledge is not only possessed individually but is also created by and shared among learners in specific contexts [13]. This notion implies that, on the one hand, while assessing the impact of LADs, sociocultural researchers could focus on their ability to support students' social and communicative learning processes (e.g., the gradual uptake of scientific vocabulary, productive conversations, active participation) rather than the intrinsic capability of individual students (e.g., based on exam scores) [13]. For example, [14] measured the impact of a LAD based on students' participation in chat discussions. On the other hand, researchers taking a cognitivist perspective tend to assess the impact of LADs using students' grades/academic achievement, which is used as a proxy for identifying students' mental functioning. [7] used a LAD to inform students about their study progress and chances of success by showing visual performance indicators of the student and the average of the cohort. The researchers measured the impact of the LAD by assessing student performance in the final programming exam.

In terms of one widely cited model of SRL [15], the impact of LADs tends to be measured along the three loosely sequenced phases of SRL (e.g., forethought, performance, reflection). In this regard, other than only looking at the content of learners' actions, SRL researchers evaluate how LADs could support students in becoming better regulators of their learning process through actions such as improved time management and goal-setting skills, which could all be linked to improved learning outcomes. For example, [16] developed NoteMyProgress, an SRL LAD that helps students keep track of their activity and create awareness of their SRL strategies in a MOOC course. [17] longitudinal study used a mobile app with visualisations showing students' learning time. The study showed that the mobile dashboard improved students' awareness of time management, which is a key SRL skill. Applying these indicators based on theoretical premises and practical 'considerations' can enable an understanding of the impact of LADs on students' learning outcomes. In this paper, we do not rely on specific theoretical perspectives or criteria to evaluate the impact of LADs; instead, we take a bottom-up approach to extract quantifiable measures of students' learning outcomes as reported by individual studies.

## 2.2 Existing reviews on LADs

[2] conducted one of the earliest reviews on LADs. Based on the analysis of 55 studies, the review found that the impact of LADs was hard to establish given the early stages of the field and suggested the need for comparative and longitudinal evaluations to establish impact. [6] reviewed 93 studies on student-facing LADs and educational recommender systems. The review was exclusively focused on LA systems that collect click-level student data and report this data directly to students. This study emphasizes the need for experiments to determine the effect of these systems on student behaviour, achievement, and skills.

[18] conducted a systematic literature review on LADs. They found that LADs lacked a strong theoretical basis, metacognitive support, insight into effective learning tactics, and robust evaluation. Although this study highlighted the evaluation of the impact of LADs on learning and teaching as one of the goals, this was not analysed and reported. [19] conducted a systematic literature review on 56 studies applying LA to enhance SRL. Findings showed that only 46% of the studies demonstrated a positive impact on learning, while only four studies reported positive effects across all three SRL phases (planning, performance, and reflection). While this study attempted to analyse the effect of LA interventions, the focus was only on SRL. Moreover, due to the limited number of studies conducted in similar settings, the authors could not conduct a proper meta-analysis but instead qualitatively reported the effect sizes reported by the studies. More recently, [4] conducted a systematic literature review on 50 teacher-facing LADs in higher education. Findings showed that most LADs raise awareness but lack actionable insights for intervention and that LAD studies pay little attention to analysing teaching and learning impact. Again, similar to previous LAD reviews, the impact of LAD on learning outcomes was not reported.

While there have been several literature reviews on LADs and provided important contributions to the area of LAD research, the available literature shows that none of these has exclusively focused on analysing the effect of LADs on students' learning outcomes. In a recent commentary published in the Journal of Learning Analytics, Ochoa argues that most existing LA studies are focused on presenting data (e.g., through LAD reports) back to stakeholders to support sense and decision-making but with little attention to evaluating the effects of the shared data on students' learning outcomes [25]. While some individual studies have reported the positive impact of LADs on students' learning behaviour and outcomes, the findings tend to be based on single studies and often small samples, which makes it difficult to generalise [20]. Yet, if we are to identify if LADs address the main goal of LA (e.g., developing insights to optimise learning and teaching), strong evidence based on the evaluation of LA interventions is needed [1]. This motivates the focus of the present study.

### 2.3 Research question

Considering the identified gaps in existing review studies, this study evaluates the impact of LADs on students' learning outcomes. The following research question guided the study: "What is the impact of LADs on students' learning outcomes?" [11] emphasised that the primary purpose of LA is to enhance learning and that the true test for LA is to show a longer-term impact on student learning and teaching practice. This research question delves into the impact of LADs on various dimensions of students' learning outcomes (e.g., achievement, participation, engagement, motivation, attitude) as reported by the analysed studies. By scrutinizing these dimensions of learning outcomes, this research provides a comprehensive understanding of how LADs can potentially transform the educational experience, offering insights into their benefits and potential challenges in contemporary learning environments.

## 3 METHODOLOGY

This study followed the Preferred Reporting Items of Systematic Review and Meta-Analysis (PRISMA) framework updated by [21] and [8]. In the following section, we describe the steps followed and the statistical approaches used to find, code and analyse the effect sizes from the included studies.

Literature Search: On the 15th of June, 2023, we searched four scientific databases relevant to our research questions: Scopus, Web of Science, ACM and ERIC. We searched for the terms (widget* OR dashboard*) AND ("learning analytics" OR "educational data mining" OR "educational datamining") in the title, abstract and author keywords. We further limited the results to journal articles, conference proceedings and book chapters published in English. The search yielded 485 articles from Scopus, 153 from Web of Science, 88 from ERIC, and 86 from ACM (See Figure 1). After removing duplicates, the remaining number of articles was 527. These results were exported to a web-based review software, Rayyan-ai [22], which was used to sort the papers. Rayyan simplifies the process by allowing users to upload citations and full-text articles within a single review. In this study, Rayyan allowed all five authors to code the included papers collaboratively and to detect and resolve inconsistencies and conflicts [22]. After sorting papers in Rayyan-ai, 57 articles remained, and 470 were excluded. The 57 included papers were exported to MsExcel for detailed coding.

Literature Filtration and Inclusion Criteria: The search results were screened according to their relevance to the topic and quality. We first excluded the results unrelated to LADs according to the titles, abstracts, and methods. Moreover, since the focus of the study was to review the effects of LADs on students' learning outcomes, we excluded theoretical and technical papers aimed at developing algorithms. The included studies had to be (1) empirical studies, (2) contain quantitative data enough for the effect size synthesis, (3) contain well-designed research procedures (e.g., separate participants into the control and experimental groups), (4) be written in English, and (5) be published in peer-reviewed journals. The final included studies amounted to 38.

Coding of included articles: Guided by the research question, we retrieved the following qualitative and quantitative data from the included studies: features of the dashboard, level of education, discipline, target audience, study design (using [23] classification), type of learning outcome indicator reported, type of evaluation, and statistical information (e.g., effect size, mean values, standard deviations, sample sizes). Initially, we were interested in conducting a meta-analysis. However, during the analysis, we found very few studies with congruent research setups and all the statistical information necessary to allow a meta-analysis. Whenever enough information was provided, we classified the variables for sub-group analysis and converted the effect size to a common unit (Cohen's d) to facilitate comparison. When using Cohen's d, a value of 0.2 indicates a small effect size; a value of 0.5 is a medium one, and a value over 0.8 is a large one [24]. For studies with insufficient information, we extracted the reported quantitative metrics (e.g., sample, effect size, mean) and reported them descriptively (see Tables 1-4). This study used a bottom-up approach to identify the learning outcome indicators reported in the analysed studies. The analysis revealed four main learning outcome indicators: (1) academic achievement, which includes an exploration of whether the use of LADs has a discernible effect on student's academic performance, including their grades, assessment scores, and overall performance in educational contexts; (2) motivation, which focuses on LADs potential to bolster students' motivation to learn; (3) participation including active involvement in classroom discussions, group projects, and other activities, and, (4) attitudes toward learning, such as self-efficacy and anxiety. These dimensions formed the basis of the analysis and the findings reported in this study.

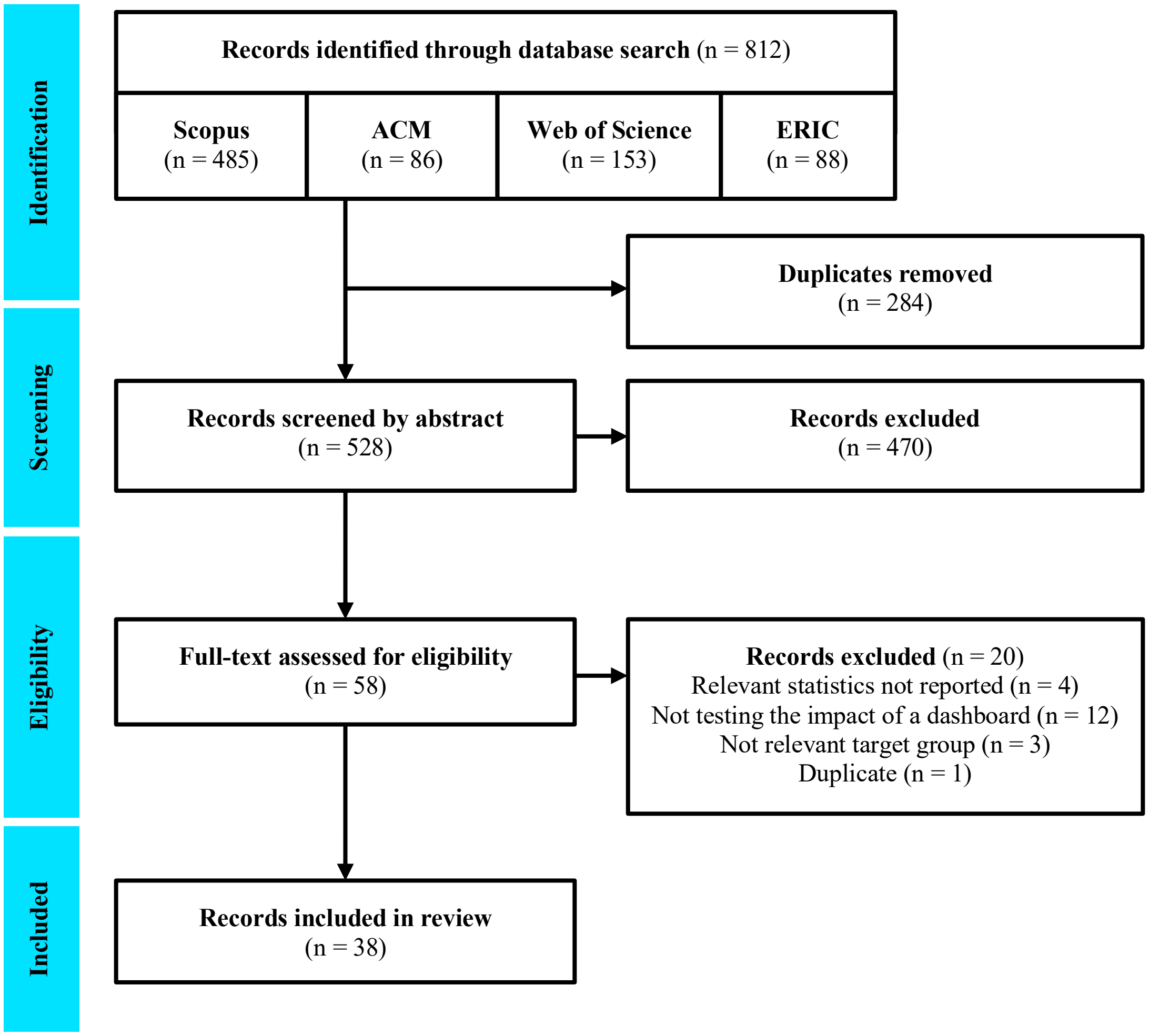

Figure 1: PRISMA flow diagram of records included in the review.

# 4 RESULTS

## 4.1 Overview of the results of the final corpus

The systematic review included a final corpus of 38 research papers (see Table AI in appendices for full details). Descriptively, we mapped the articles according to the level of education, dashboard target group, disciplines, and study design (See Figures 2 and 3). With respect to the level of education, the majority (68.4%) of the included studies focused on higher education, primarily involving participants with advanced degrees (undergraduate and graduate students). Pre-higher education studies represented a substantial portion, comprising approximately one-third (26.3%) of the selected papers, with dashboards targeting students from primary schools [26] and secondary schools [27].

The remaining studies (5.3%) fell into the "other" category, primarily emphasising professional development as their educational context, demonstrating a smaller but noteworthy presence in the corpus. In examining dashboard target groups across the 38 selected studies, the majority (78.9%) of the studies focused solely on students as the target group for educational dashboards. A smaller proportion (10.5%) of studies explored the effect size for both students and teachers. Unsurprisingly, fewer studies looked at teachers' only dashboards and those with advisors since the intention of the study was to explore dashboards reporting on students' learning outcomes.

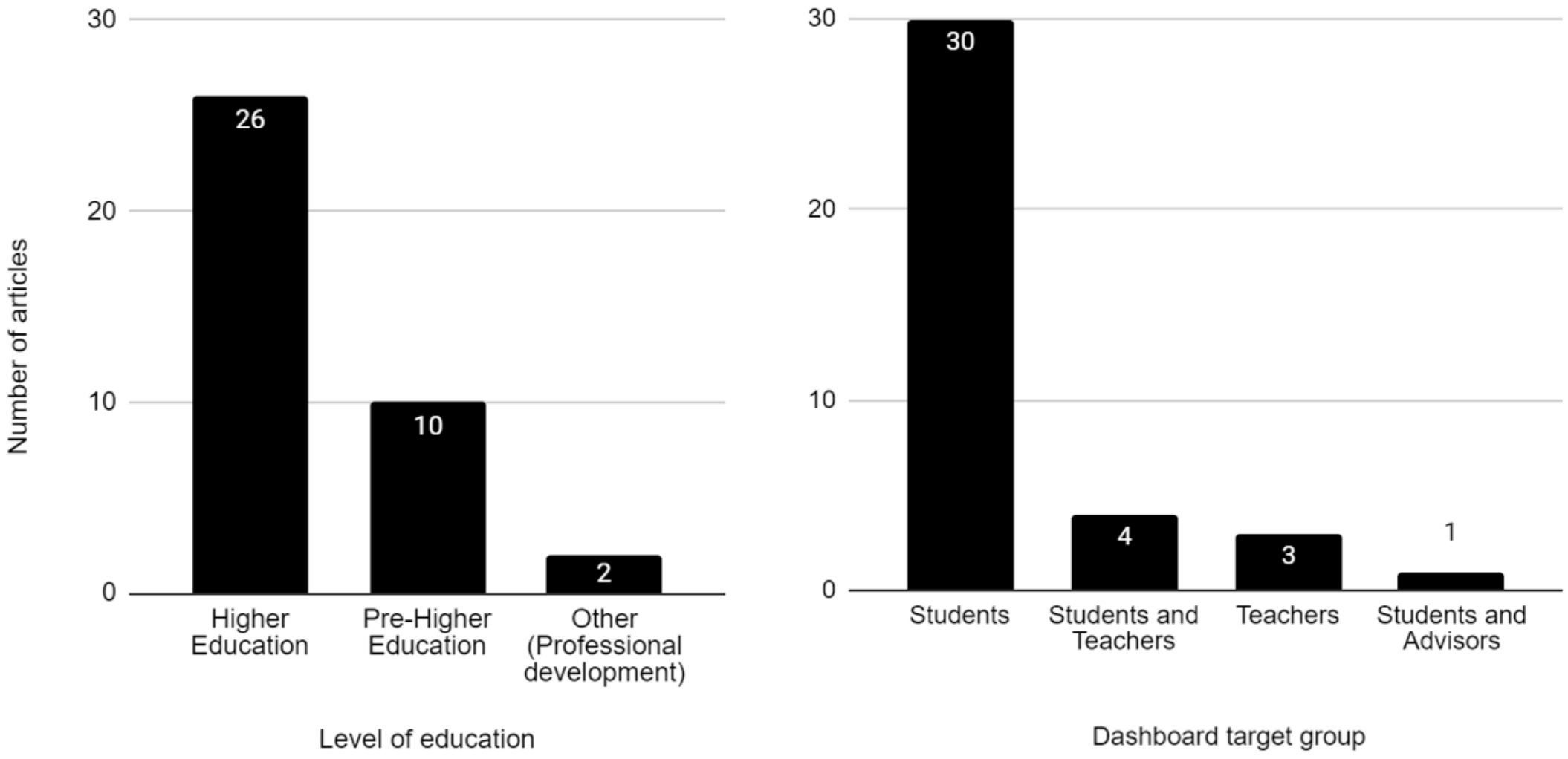

Figure 2. Included studies' level of education and dashboard target group (*n*= 38)

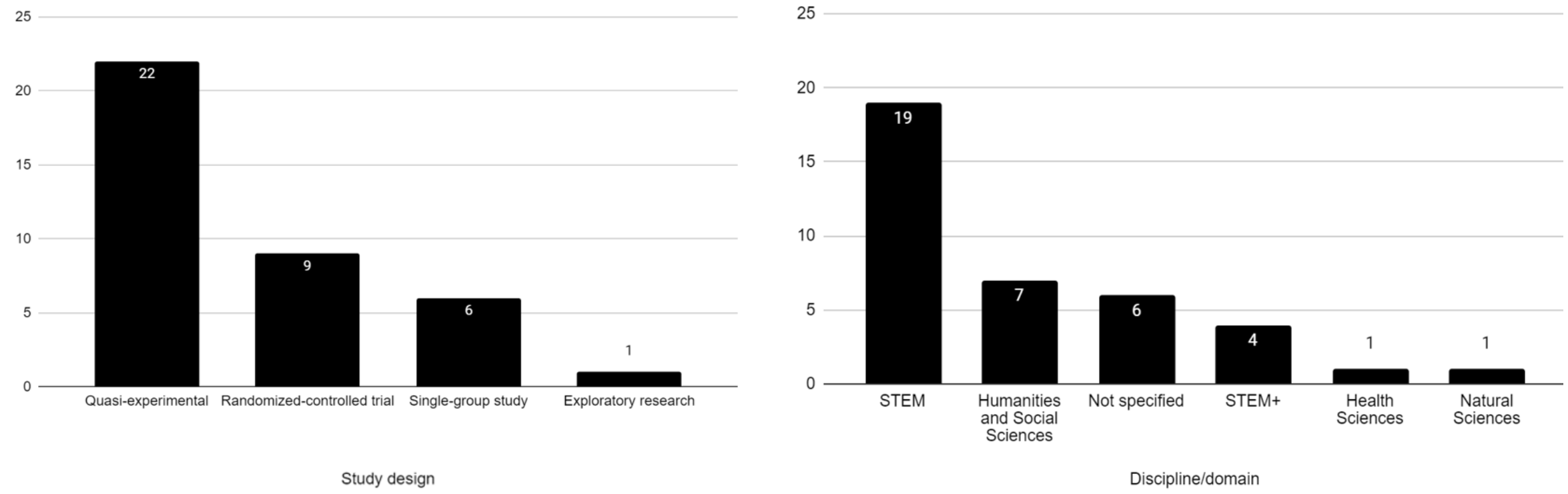

Figure 3. Included studies design and discipline/domain (*n*= 38)

In following [23] to classify the study design of the final corpus, a plurality (57.9%) adopted a quasi-experimental design in which study groups were assigned using a non-randomised method, e.g., [28] and [29]. Nine research studies (23.7%) explored randomised controlled trials (RCTs), ensuring a rigorous evaluation of interventions through random allocation (e.g., [20]. Six studies (15.8%) used a single-group design, while one adhered to an exploratory research design. The dominant context in terms of the discipline where dashboards were experimented with was in STEM (Science, Technology, Engineering, and Mathematics), as demonstrated by half (50.0%) of the included studies. Humanities and Social Sciences accounted for (15.78%) of the experiments, followed by STEM+ (10.52%). Lastly, two single studies were applied in Health Sciences and Natural Sciences, respectively.

## 4.2 Impact of LADs on students' achievement

**Table 1.** Studies reporting on the impact of LADs on students' achievement

| Study | Study design | Measure | Substudy | N | Effect size (Cohen's d) | | |
|---|---|---|---|---|---|---|---|
| Afzaal et al. (2021) | Quasi-experimental (used vs. not) | Final grade | | 40 | 2.170 | *** | Large |
| Amarasinghe et al. (2020) | RCT | Perceived learning outcomes | | 813 | 0.103 | | Negligible |
| Davis et al. (2016) | RCT | Pass rate | | 3188 | 0.093 | | Negligible |
| | | Final grade | | | n/a | * | n/a |
| Dickler et al. (2021) | Quasi-experimental | Activity grade | | 70 | 0.721 | * | Medium |
| Duan et al. (2022) | Single-group study | Grade rank | | 69 | 0.750 | ** | Medium |
| Fleur et al. (2020) | RCT | Final grade | | 72 | 0.489 | * | Small |
| Han et al. (2021) | Single-group study | Perceived learning outcomes | | 85 | 0.365 | | Small |
| Hellings et al. (2022) | RCT | Grade of the course | | 556 | 0.110 | | Negligible |
| | | Exercise grade | | | 0.482 | *** | Small |
| | | Passing rate | | | 0.314 | *** | Small |
| | | Quiz grade | | | 0.247 | *** | Small |
| Herodotou et al. (2019) | Quasi-experimental | Final grade | | 559 | 0.223 | * | Small |
| | | Pass rate | | | 0.173 | *** | Negligible |
| Jonathan et al. (2017) | Single-group study | Critical reading fluency | | 101 | 0.220 | ** | Small |
| | | Critical reading ability | | | 0.320 | ** | Small |
| Kim et al. (2016) | Experimental | Score after midterm | | 151 | 0.420 | * | Small |
| Kokoç et al. (2021) | Exploratory research | Final grade | | 126 | 2.787 | *** | Large |
| Manganello et al. (2021) | Quasi-experimental (used vs. not) | Pre-test/Post-test | | 41 | 0.770 | * | Medium |
| Molenaar et al. (2017) | Quasi-experimental (used vs. not) | Arithmetic skills | | 1472 | 1.790 | *** | Large |
| Park et al. (2015) | RCT | Final exam | Class A | 43 | 0.065 | | Negligible |
| | | | Class B | 30 | 0.036 | | Negligible |
| Tlili et al. (2021) | RCT | Pre-test/Post-test | | 31 | 2.336 | *** | Large |
| Toyokawa et al. (2023) | Quasi-experimental | Content | | 37 | 0.100 | | Negligible |
| | | Language use | | 37 | 0.192 | | Negligible |
| | | Vocabulary | | 42 | -0.487 | | Small |
| Valle et al. (2021a) | RCT | Final exam | Descriptive | 125 | 0.278 | | Small |
| | | | Predictive | 123 | 0.138 | | Negligible |
| | | Quiz grade | Descriptive | 125 | 0.193 | | Negligible |
| | | | Predictive | 123 | 0.052 | | Negligible |
| Valle et al. (2021b) | RCT | Quiz grade | | 146 | 0.002 | | Negligible |
| | | Final exam | | | 0.017 | | Negligible |
| Xhakaj et al. (2017) | RCT | Pre-test/Post-test | | 300 | 0.209 | | Small |
| Yang et al. (2022) | RCT | Reading speed | | 47 | -0.006 | | Negligible |
| Zheng et al. (2022) | Quasi-experimental | Collaborative knowledge building | | 60 | 2.168 | *** | Large |

Most of the studies in our review focusing on the impact of LADs on students' performance had either negligible (i.e., marginal), 14 studies (41.2%), or small effect size, 12 studies (35.3%), totalling 26 studies (76.5%) (See Table 1). Three studies (8.8%) had medium effect sizes; in two of the studies, [30] and [31] compared students who used the dashboard to those who did not. A comparison that essentially measures the difference between two activity levels (not a comparison between a control and an experimental group). The third study by [32] used the dashboard to alert the teacher to help the students. As such, the effect size is mostly a measure of the magnitude of the effect of teacher intervention. A total of five studies (14.7%) reported large effect sizes; of these, two studies, [33] and [34], compared activity levels of usage. Another study by [26] compared an LA-enabled dashboard to another control group that used paper and pen. The LAD arm also benefited from learning with technology, teacher instructions and paced practice. The remaining two studies reporting large effect sizes assessed the effects of the dashboard on a collaborative activity. In [35], the dashboard arms were found to be able to name animals more than the control arms. In [36], the difference was evaluated through teachers' ratings of collaborative knowledge-building maps built by the teachers.

## 4.3 Impact of LADs on students' participation

**Table 2.** Studies reporting on the impact of LADs on students' participation

| Study | Study design | Measure | N | Effect size (Cohen's d) | |
|---|---|---|---|---|---|
| Aguilar et al. (2021) | Single-group study | Effort and Persistence | 202 | -0.295 | Small |

| | | | | | | |
|---|---|---|---|---|---|---|
| Aljohani et al. (2019) | Quasi-experimental | Access to LMS | 86 | 0.821 | *** | Large |
| | | Number of discussion threads | | 1.379 | *** | Large |
| | | Access to discussion board | | 0.916 | *** | Large |
| Amarasinghe et al. (2020) | Quasi-experimental | Activity participation | 813 | 0.532 | *** | Medium |
| | | Discussion participation | | 0.437 | *** | Small |
| | | Voting participation | | 0.379 | *** | Small |
| Davis et al. (2016) | RCT | Quiz submissions | 3188 | 0.084 | * | Negligible |
| Han et al. (2021) | Single-group study | Number of comments | 88 | 1.392 | *** | Large |
| | | Participation | | 0.715 | ** | Medium |
| | | Interaction | | 0.806 | *** | Large |
| Harvey et al. (2020) | Quasi-experimental (different usage levels) | Persistence (time on task) | 238 | 0.208 | | Small |
| Hellings et al. (2022) | RCT | Number of passing assignments submitted | 556 | 0.180 | ** | Negligible |
| Jonathan et al. (2017) | RCT | Cognitive reading engagement | 101 | 0.559 | *** | Medium |
| Yang et al. (2022) | RCT | Book count | 47 | 0.655 | ** | Medium |
| | | Time read (min) | | 0.649 | * | Medium |
| | | Unique days | | 0.890 | * | Large |

Out of the 38 studies included in this review, nine of them studied the impact of LADs on students' participation (see Table 2). Four studies reported either small statistically insignificant or negligible statistically significant effects [37, 38, 40], while five studies reported significantly large or medium effect size. One of these studies is the one by [41], who employed a quasi-experimental design, where the experimental group was exposed to a dashboard which presented their Learning Management System (LMS) participation patterns compared to the most engaged students and the class average. The results revealed large significant effect sizes across multiple measures of student engagement, such as access to the LMS (d=0.821), number of discussion threads (d=0.379), and access to discussion boards (d=0.916). [42] introduced a dashboard for adaptive support in face-to-face collaborative argumentation that visualised three main indicators: opinion distribution, participation and interaction, as well as the use of argumentation elements, such as claims, reasons or originality. This single-group study reported a medium effect size (d=0.715) on peer participation. The quasi-experimental study by [43] examined the impact of a dashboard visualising the reading outcomes, the level of self-directed learning skills, and feedback to improve students' skill levels. The results showed that using the dashboard had a large significant effect on the number of unique reading days (d=0.890) and medium significant effects on the number of books viewed (d=0.655) and on the number of minutes spent in the reading system (d=0.649). [44] conducted an RCT using a dashboard with students' own data as a reference frame visualising, among others, the frequency and length of student comments, comment types, and a social network of student interactions. A statistically significant medium effect on cognitive reading engagement was reported (d=0.559). Finally, a single-group study by [45] examined the effect of teachers using a dashboard to orchestrate scripted collaborative learning sessions on student collaboration. The dashboard not only visualized indicators of student collaboration but also enabled teacher interventions. The results showed that teachers using a dashboard had a medium effect on students' activity participation (d=0.523)

## 4.4 Impact of LADs on students' motivation

A total of six out of the 38 included studies investigated the impact of dashboards on different motivation-related constructs. As illustrated in Table 3, the effect size across the studies ranged between negligible [39, 43], small [46], medium [35], and large [48]. [48] conducted an experimental study where participants were randomly assigned to either the treatment group (access to a dashboard, n=34) or the control group (no access, n=38). The authors used two measures of motivation: intrinsic and extrinsic. Based on a survey and assessment scores, the findings showed that the treatment group showed a large effect size (d=0.809) on extrinsic motivation at the end of the course compared to the control group.

**Table 3.** Studies reporting on the impact of LADs on students' motivation

| Study | Study design | Measure | Substudy | N | Effect size (Cohen's d) | | |
|---|---|---|---|---|---|---|---|
| Aguilar et al. (2021) | Single-group study | Instrumental motivation | | 176 | -0.129 | | Negligible |
| Fleur et al. (2020) | RCT | Intrinsic motivation | | 26 | -0.042 | | Negligible |
| | | Extrinsic motivation | | | 0.809 | | Large |
| Tlili et al. (2021) | RCT | Motivation | | 31 | 0.795 | ** | Medium |
| Valle et al. (2021a) | RCT | Intrinsic goal orientation | Descriptive | 125 | 0.234 | | Small |
| | | | Predictive | 123 | -0.157 | | Negligible |
| | | Extrinsic goal orientation | Descriptive | 125 | 0.137 | | Negligible |
| | | | Predictive | 123 | -0.351 | * | Small |
| Valle et al. (2021b) | RCT | Intrinsic goal | | 146 | -0.060 | | Negligible |
| | | Extrinsic goal | | | -0.338 | * | Small |
| | | Task value | | | -0.023 | | Negligible |
| | | Control beliefs | | | -0.243 | | Small |

| | | | | | | |
|---|---|---|---|---|---|---|
| | | Self-efficacy | | | -0.260 | Small |
| | | Test anxiety | | | -0.145 | Negligible |
| Yang et al. (2022) | RCT | Learning motivation | | 47 | -0.011 | Negligible |

MSLQ = Motivation Strategies for Learning Questionnaire

## 4.5 Impact of LADs on students' attitudes

A total of eight out of the 38 studies included have investigated the impact of dashboards on a wide variety of constructs related to learning attitudes (see Table 4 and Table A2 in appendices). Most studies have reported non-significant, negligible, or small effect sizes, with few exceptions. [22] used a LAD for facilitating collaborative argumentation in face-to-face learning and found a medium impact (d=0.609) of the dashboard on students' situational interest and a large one (d=1.727) on group regulation using a single-group study design (before and after using the dashboard). [40] investigated how graphs of students' relative engagement (compared to their classmates) —presented on a dashboard— affected their self-esteem and found small to medium differences (d=0.414) depending on whether the students were shown positive or negative information. [36] found that students who received personalised feedback during a collaborative learning task had greater positive emotions (d=0.820) than students who did not, while the difference in neutral and negative emotions was insignificant. Notable also is the study of [47], who found that a descriptive LAD decreased students' fear and anxiety with a small (yet significant) effect size, whereas a predictive dashboard did not yield any significant effect.

**Table 4.** Studies reporting on the impact of LADs on students' attitudes

| Study | Study design | Measure | Substudy | N | Effect size |
|---|---|---|---|---|---|
| Aguilar et al. (2021) | Single-group study | PALS, SRLQ | | 201 | Negligible - Small |
| Aslan et al. (2019) | Quasi-experimental | Confusion, Satisfaction, Boredom | | 37 | Non-significant |
| Han et al. (2021) | Single-group study | Situational interest & Group regulation | | 85 | Medium - Large |
| Harvey et al. (2020) | Quasi-experimental (different usage groups) | Self-esteem | | 238 | Small - Medium |
| Valle et al. (2021a) | RCT | MSLQ, STARS | Descriptive | 125 | Negligible - Small |
| | | | Predictive | 123 | Non-significant |

| | | | | |
|---|---|---|---|---|
| Valle et al. (2021b) | RCT | MSLQ, STARS | 146 | Negligible - Small |
| Wang et al. (2023) | RCT | Learning attitude | 145 | Non-significant |
| Zheng et al. (2022) | RCT | Positive emotion | 60 | Large |

PALS = Patterns of Adaptive Learning Scale; SRLQ = Self-Regulated Learning Questionnaire; MSLQ = Motivation Strategies for Learning Questionnaire; STARS = Statistical Anxiety Rating Scale.

# 5 DISCUSSION

This review, based on 38 studies, sought to systematically assess the impact of LADs on different aspects of students' learning (i.e., achievement, participation, motivation, and attitude). Given the differences between statistical methods, we standardised – and computed –all effect sizes for comparability. Our analysis has shown a vast landscape of varying impacts of LADs on academic performance. Yet, the overarching conclusion is that LADs have not lived up to the promise and most of the studies reported either negligible or small impact on students' performance, suggesting that the implementation of LADs alone may not significantly enhance performance as once hoped. In almost all the studies that reported medium or large effect sizes, though few, LADs have either been combined with another type of intervention (and thus had an obvious confounding) or were assessed using a non-controlled design. It is also worth mentioning that our systematic review did not include a single well-powered controlled study that assessed a LAD with properly randomised samples. Such findings show that the methodological design and reporting methods of the studies in our dataset did not allow a rigorous assessment. This was especially the case when researchers compared users and non-users of LADs. For example, [30] and [31] reported a medium effect size of the impact on achievement, but their designs primarily compared the activity levels of the dashboard i.e., users vs. non-users, which is neither randomised nor equal to any form of intervention vs. control. Users of the dashboard are, in this case, engaged students who utilise the learning resources and the tools that help them learn. Non-users are probably students who are less willing to engage or consume the learning resources. At the same time, even when studies use a rigorous experimental design where one group had access to a dashboard and one group did not, it is important to note that within the group that had access, not all students may have actually used the dashboard. As such, the comparison is between those who had the opportunity to use it (access) versus those who did not, regardless of whether everyone in the access group actively used it or not. Therefore, future research should prioritise methodological rigour by employing better comparative designs that help disentangle the dashboard effect from other confounding factors. Modern statistical techniques such as propensity score matching, propensity score weighting, and regression discontinuity designs can help address issues related to self-selection bias and provide more robust and accurate assessments of the impact of LADs [42].

The pool of the available studies kept us pondering the study design, metrics, and evaluation of LADs. In that, all studies that reported large effect sizes were underpowered with small sample sizes except [47] which had a sample size of 373; studies with small or no effect size had sample sizes of 3188 (e.g., [37] and 813 (e.g., [45]). We also observed that the potential impact of LADs on students' learning outcomes may vary depending on the dashboard features and the specific learning activities they support. For example, [46] used two different types of dashboards as an intervention —a descriptive and a predictive dashboard. The study found that a predictive dashboard significantly reduced learners' interpretation anxiety and had an effect on intrinsic goal orientation as compared to the descriptive dashboard. At the same time, the study reported no effect on students' performance measured based on the final exam and total quiz scores. This implies that different dashboard designs and measures might have different learning outcome impacts. Future research should focus on uncovering strategies and conditions that lead to substantial improvements in learning outcomes. This could involve investigating specific dashboard design features, instructional interventions, or contextual factors that have the potential to generate large effects.

Even though the analysed studies are about LA, the methods used in the evaluation of the dashboards were very traditional. Future work would benefit from using LA methods to identify students' profiles of usage of LADs and to analyse how they are used within the learning process. The analysis revealed a predominant reliance on surveys, online quizzes, and crude proxies such as course grades and teacher-scored rubrics as means to evaluate the impact of LADs. For example, in the study by [36], the difference between the control and experimental groups was evaluated through teachers' ratings of collaborative knowledge-building maps built by the teachers. While such methods can provide insight into the impact of a LAD intervention, these are prone to subjectivity bias and errors, which could limit an accurate analysis of the impact of LA dashboards. Moreover, with a few exceptions (e.g., [35, 30], few studies used pre-and post-test exams to assess the impact of the LADs, which makes it difficult to properly assess the actual impact of the dashboards before and after the interventions. To advance the field, we believe that researchers should explore more comprehensive and nuanced assessment approaches that accurately capture learning processes and products.

The analysis revealed a limited availability of standards for evaluating learning constructs like attitudes in the context of LADs, which poses a challenge to research consistency and comparability. Researchers should work towards establishing clear and validated standards for assessing specific constructs related to learning outcomes, such as attitudes, motivation, and engagement. Developing standardised measurement tools and evaluation criteria will enhance the reliability and validity of findings across studies, allowing for more meaningful comparisons and generalisations.

# 6 LIMITATIONS

This study offers valuable insights into the impact of LADs on various aspects of students' learning outcomes. However, several limitations should be considered when interpreting the findings and implications. Firstly, a notable limitation is the inability to conduct a meta-analysis as initially anticipated. This limitation arose due to missing information in the analysed studies, the different design and confounding variables. Future research in the field could benefit significantly from standardised reporting practices that include all relevant details to facilitate meta-analyses. Secondly, our analysis focused solely on the impact of LADs on students' learning outcomes. Dashboards designed to improve teaching and instructors' outcomes were not within the scope of this study. Future research could explore the effects of these instructor-oriented dashboards, considering their potential to enhance teaching strategies and support educators in optimising the learning experience. Moreover, additional variables such as differences in dashboard implementation, disciplines, and theoretical perspectives and how they impact the effect of dashboards require exploration. Despite these limitations, this study serves as a valuable stepping stone in understanding the multifaceted impact of LADs on students' learning outcomes. To address the observed limitations, future research endeavours aim to extend the study to a meta-analysis. This extension will become feasible with the acquisition of additional information from authors, emphasising the importance of comprehensive reporting in research publications.

# 7 CONCLUSION AND IMPLICATIONS FOR FUTURE STUDIES

Across the spectrum of performance/achievement, participation, motivation, and attitude, our analysis revealed a vast landscape with varying impacts of LADs on students’ learning outcomes. Performance outcomes, often regarded as a key measure of success, showed predominantly negligible or small effects, with a dearth of robust controlled studies to establish causal relationships. As we currently stand, there is no evidence to support the conclusion that LADs can improve academic achievement. Similarly, motivation and attitude improvements were generally modest, though notable exceptions underscored the potential of LADs to influence these constructs positively in specific contexts. Participation emerged as the area where LADs demonstrated a relatively average impact, with some studies reporting medium to large effect sizes. This suggests that LADs may hold promise in enhancing student engagement and interaction in online learning environments. However, our analysis also underscores the methodological limitations in the existing body of research, including small sample sizes, reliance on traditional evaluation methods, and a lack of standardised assessment tools that raised concerns about the accuracy of measurements and the potential oversight of deeper learning processes. Diversifying assessment methods to capture a more comprehensive view of learning outcomes is imperative for advancing the field. Future investigations should prioritise rigorous research designs, explore nuanced evaluation approaches, and establish clear standards for assessing learning constructs related to LADs. By doing so, we can move closer to harnessing the full potential —if it at all exists— of LADs in optimising teaching and learning through data-informed decision-making for educators and learners.

## 8 ACKNOWLEDGEMENTS

R.K was funded by the Norwegian Research Council within the project TEAMLEARN. M.S. was funded by the Academy of Finland within the project TOPEILA, Decision Number 350560 which was received by the last author. S.L.-P. and K.M. were partially funded by the European Commission within the project ISILA (2023-1-FI01-KA220-HED-000159757). M.K was partially funded by the European Commission Erasmus+ project RIALHE (2022-1-NO01-KA220-HED-000087273)